\documentclass[preprint,showpacs,aps,prb]{revtex4}

\usepackage{epsfig}

\begin{document}
\newcommand{\vektor}[1]{\mbox{\boldmath $#1$}}
\title{Competing interactions of spin and lattice in the Kondo lattice model}
\author{M. Gulacsi${}^{\rm 1}$, A. Bussmann-Holder${}^{\rm 2}$ 
and A. R. Bishop${}^{\rm 3}$}
\affiliation{
${}^{\rm 1}$ 
Department of Theoretical Physics, 
Institute of Advanced Studies \\
The Australian National University, Canberra, ACT 0200, Australia \\
${}^{\rm 2}$ 
Max-Planck-Institut f\"ur Festk\"orperforschung \\
Heisenbergstr. 1, 70569 Stuttgart, Germany \\
${}^{\rm 3}$ 
Theoretical Division, Los Alamos National Laboratory \\
Los Alamos, NM 87545, U.S.A.}

\date{20 June 2003}

\begin{abstract}
The magnetic properties of a system of coexisting localized 
spins and conduction electrons are investigated within an 
extended version of the one dimensional Kondo lattice model
in which effects stemming from the electron-lattice and  
on-site Coulomb interactions are explicitly included. After 
bosonizing the conduction electrons, is it observed that 
intrinsic inhomogeneities with the statistical scaling 
properties of a Griffiths phase appear, and determine 
the spin structure of the localized impurities. The 
appearance of the inhomogeneities is enhanced by appropriate
phonons and acts destructively on the spin ordering. The 
inhomogeneities appear on well defined length scales, 
can be compared to the formation of intrinsic mesoscopic 
metastable patterns which are found in two-fluid
systems. 
\end{abstract}

\maketitle

The interplay of spin, charge and lattice degrees of freedom 
has been investigated intensively in many transition metal 
oxides and especially in perovskite manganites, which 
have recently attracted new interest due to the discovery of 
colossal magnetoresistance (CMR). The initial understanding
of the properties of manganites was based on the double-exchange
(DE) mechanism within the Kondo lattice.\cite{deCMR} However, 
the new experimental findings \cite{phonontheory} have revealed 
that this approach is incomplete and has to be extended to 
account for effects stemming from the lattice in order to 
understand the doping dependent phase diagram and the 
richness of phases that appear. In the following we model 
these complex systems within the Kondo lattice model (KLM), 
admitting for ferro- and antiferromagnetic couplings, and 
including explicitly the interaction with the lattice degrees 
of freedom.

The KLM considers the coupling between half-filled narrow 
band (localized $d$ or $f$) and conduction electrons. Even 
though studied intensively for the last two decades, 
the understanding of the KLM remains incomplete. Only in 
one dimension have numerical simulations \cite{numerics} and 
bosonization techniques \cite{boso,graeme} have been 
carried through which admit predictions about the phase 
diagram of the KLM. No investigations exist for the case 
where the KLM is extended to account for contributions 
stemming from the phonons, which is of special relevance 
to CMR materials. In particular, the small doping regime of these 
systems, which are ferromagnetic at low temperatures, 
seems to be appropriate to be modeled within the KLM 
extended by interactions with the lattice. In the 
following we present bosonized solutions of the KLM 
where on-site Coulomb and phonon contributions are 
explicitly included. This ``extended'' KLM model allows 
spin-and magnetoelastic-polaron formation, which we believe are 
of major importance in understanding these complex materials. 

The Hamiltonian of the KLM in the presence of on-site Coulomb 
interaction reads: 
\begin{eqnarray}
H_{\rm KLM} \: &=& \: -t \sum_{j, \sigma} (c^{\dagger}_{j, \sigma} 
c^{}_{j+1, \sigma} + {\rm{h.c.}}) 
\nonumber \\
&+& \: J \sum_{j} {\bf S}_{{\rm d}, j} 
{\bf \cdot} {\bf S}_{{\rm c}, j} + U \sum_{j} n_{j, \uparrow} 
n_{i, \downarrow} \; , 
\label{one}
\end{eqnarray}
where $t > 0$ is the conduction electron hopping integral,
${\bf S}_{{\rm d}, j} = \frac{1}{2} \sum_{\sigma^{},\sigma^{\prime}}
c^{\dagger}_{{\rm d}, j, \sigma^{}} {\vektor{\sigma}}_{\sigma^{}, 
\sigma^{\prime}} c^{}_{{\rm d}, j, \sigma^{\prime}}$, 
${\bf S}_{c, j} = \frac{1}{2} \sum_{\sigma^{},\sigma^{\prime}} 
c^{\dagger}_{j, \sigma^{}} {\vektor{\sigma}}_{\sigma^{},
\sigma^{\prime}} c^{}_{j, \sigma^{\prime}}$ and 
${\vektor{\sigma}}$ are the Pauli spin matrices.  Fermi 
operators $c^{}_{j, \sigma}, c^{\dagger}_{j, \sigma}$ with 
subscript $d$ refer to localized $d$-spins, while those not 
indexed refer to the conduction electrons. The on-site Coulomb 
repulsion is given by the Hubbard term proportional to $U$. 
In the CMR materials the localized states are represented by 
the threefold degenerate Mn $t_{2g}$ $d$-electrons with total 
spin 3/2. However, for reasons of transparency, the localized spin 
is approximated here by spin 1/2, since the properties of the model 
are qualitatively independent of the magnitude of the localized spins. 
In the following the the Kondo coupling $J$ is measured in units 
of the hopping $t$ and both cases, antiferromagnetic ($J > 0$) 
and ferromagnetic ($J < 0$) couplings, will be considered. The 
conduction band filling is given by $n = N_{c}/N < 1$, where $N$ 
is the number of lattice sites and $N_{c}$ is the number of
conduction electrons. To be able to understand the wide
range of properties of the CMR materials, we also allow 
for the number of impurity spins,  $N_{d}$, to vary, 
in such a way that $N_{d} / N < 1$. 

In principle, the electron-phonon coupling could be of either 
inter-site (Su-Schrieffer-Heeger (SSH) \cite{ssh}) or on-site
(Holstein \cite{holstein}) character. Since we found the 
SSH-coupling to be irrelevant to forward scattering processes, 
its influence will not be discussed in the following; 
only terms arising from the on-site couplings,\cite{holstein}
i.e., $\sum_{j} \alpha q^{}_{j} n^{}_{j}$, with coupling
constant $\alpha$ and displacement $q_{j}$ will be included. 
The bare lattice Hamiltonian is: $H_{\rm latt.} = 
\sum_{j} [ p^{2}_{j} / 2 M + K q^{2}_{j} / 2$, 
where $p_{j}$ is $q_{j}$'s conjugate momenta, $K$ the harmonic 
coupling and $M$ the ionic mass. 

The underlying bosonization scheme follows standard 
procedures \cite{haldane} by first decomposing the 
on-site operators into Dirac fields, 
$c^{}_{x, \sigma} \: \approx \: 
\sum_{\tau} e^{i k_{F} x} \Psi_{\tau, \sigma} (x)$, 
where $k_{F} = \pi n /2$, 
with spinor components $\tau = \pm$ 
(+/- being the right/left movers) and $k_{F} = \pi n / 2$. 
Next we bosonize the Dirac fields with $\Psi_{\tau, \sigma} 
= \exp (i \Phi_{\tau, \sigma}) 
/ \sqrt{2 \pi \lambda}$, where $1 / \lambda$ is the ultraviolet 
cutoff. For the scalar Bose fields, $\Phi_{\tau, \sigma} (x)$,  
and their conjugate momenta, $\Pi_{\tau, \sigma} (x)$, 
$\Phi_{\tau, \sigma} (x) = \int^{x}_{-\infty} d x^{\prime} 
\Pi_{\tau, \sigma} (x^{\prime})$, are used in standard 
Mandelstam representation by means of which a momentum 
cutoff via the Fourier transform is introduced
$\Lambda (k) = \exp ( - \lambda \vert k \vert / 2 )$. 
If the distance between the impurity spins is larger than
$\lambda$, the electrons will behave as collective 
density fluctuations.\cite{haldane} Thus, the 
Fermi fields can be represented in terms of density
operators which satisfy Bose commutation relations: 
$c^{}_{\tau, x, \sigma} \approx \exp ( i \tau k_{F} x) 
\exp i \{ \theta_{\rho}(x) + \tau \phi_{\rho}(x) 
+ \sigma [ \theta_{\sigma}(x) + \tau \phi_{\sigma}(x)] \} / 2$, 
where the Bose fields for $\nu = \rho, \sigma$ are defined by
$\psi_{\nu}(x)  = i (\pi / N) \sum_{k \ne 0} e^{i k x}
[ \nu_{+}(k) \pm \nu_{-}(k) ] \Lambda(k) / k$, with $+$ 
corresponding to the number fields $\psi_{\nu} = \phi_{\nu}$ 
and $-$ to the current fields $\psi_{\nu} = \theta_{\nu}$.
The charge (holon) and spin (spinon) number fluctuations are 
defined as $\rho_{\tau}(k) = \sum_{\sigma} \rho_{\tau, \sigma}(k)$, 
and $\sigma_{\tau}(k) = \sum_{\sigma} \sigma \rho_{\tau, \sigma}(k)$. 
All rapidly oscillating terms originating from e.g. backscattering 
and umklapp processes are neglected, since they contribute only at 
exactly half filling. 

The localized $d$ electrons can neither be bosonized nor 
Jordan-Wigner transformed since no direct interaction exists 
between them. Using the continuum approximation 
for the phonon contribution, two components are
relevant: a small momentum part $\Phi_{0} (p)$ and a 
rapidly oscillating term at $2 k_F$, $\Phi_{\pi} (x)$, resulting
from the splitting of the conduction band electrons into right 
and left movers. While the former contribution causes forward 
scattering and is best represented in momentum space, the latter 
one gives rise to backscattering and requires representation in 
real space. The transformed Hamiltonian thus becomes: 
\begin{eqnarray}
H  \: &=& \: H^{\rm el} \: + \: H^{\rm ph} \: + \: 
H^{\rm el-ph} \: + \: 
\frac{J}{2 \pi}\sum_{j} [\partial_{x}\phi_{\sigma}(j)] 
S_{{\rm d}, j}^{z}
\nonumber \\
&+& \: \frac{J}{4 \pi \lambda} \sum_{j} \left\{
\cos [\phi_{\sigma}(j)] + \cos[2k_{F}j + \phi_{\rho}(j)] 
\right\} \left(e^{-i \theta_{\sigma}(j)} S_{{\rm d}, j}^{+} 
+ {\rm h.c.} \right)
\nonumber \\
&-& \: \frac{J}{4 \pi \lambda} \sum_{j} \: \sin[\phi_{\sigma}(j)]
\sin[2k_{F}j + \phi_{\rho}(j)] S_{{\rm d}, j}^{z} \; .
\label{two}
\end{eqnarray} 
If holes are present in the array of $d$-spins, all terms 
proportional to $S$ are zero. The notations used in 
Eq.\ (\ref{two}) are: the forward scattering Holstein 
electron-phonon coupling term $H^{\rm el-ph} = (\alpha / {\sqrt{M}}) 
({{\sqrt2}} / N) \sum_{p} [ \rho_{+} (-p) + \rho_{-} (-p) ] 
\Phi_{0} (p)$; the bare lattice contribution   
$H^{\rm ph} = (1 / 2 N) \sum_{p} [ \Pi^{2}_{0} (p) + 
\omega^{2}_{0} \Phi^{2}_{0} (p) ]
+ \frac{1}{2} \int dx [ \Pi^{2}_{\pi} (x) + 
\omega^{2}_{\pi} \Phi^{2}_{\pi} (x) ]$ with   
$\omega_{0} = \omega_{\pi} = \sqrt{ K / M}$; 
and the standard spinon-holon term  
$H^{\rm el} = ( v_{\rho} / 4 \pi) \sum_{j, \nu}
\{ \Pi_{\nu}^{2}(j) + [\partial_{x} \phi_{\nu}(j)]^{2} \}$ 
with velocities $v_{\rho \: / \: \sigma} \: = \: v_{F} [1 \: 
\pm \: U/ \pi v_{F} \: \mp \: \alpha^2 / \pi K v_{F}]^{1/2}$. 

It is important to note that a renormalization of the 
spinon-holon velocities appears here due to the Hubbard 
and phonon terms which act oppositely on the corresponding 
velocities. While the Hubbard term leads to a localization 
of the spinons and an increased hopping of the holons, thus 
supporting a magnetic ground state, the phonons delocalize 
the spins, but localize the charges and act destructively 
on the magnetic properties. It is worth mentioning that the 
Hubbard term alone already suffices to establish two time 
scales for the holon-spinon dynamics, but an important 
renormalization of the critical properties of the system 
is achieved through the variable phonon coupling, which Ð 
as will be shown below Ð establishes the existence of a 
Griffiths phase. The competition between the Hubbard and 
the phonon term obviously vanishes for $U = \alpha^2 / K$.  

In the following effects arising from the localized spin 
$d$ impurities, double exchange (DE), the phonons and Hubbard 
interactions will be discussed in more detail. The localized 
spin $d$ impurities act via double exchange (DE) on the hopping 
electrons so as to preserve their spin when moving through the 
lattice in order to screen the localized spins which are in 
excess of the conduction electrons, i.e. $N > N_{c}$. This,  
in turn, leads to a tendency to align the localized spins 
and results in an additional screening energy for the 
conduction electrons. 

In order to gain a more transparent understanding of this 
complicated interplay, the model is investigated first for
the case of two sites and one conduction electron,\cite{detheory} 
next in a simple continuum approximation, and finally the full 
bosonized solution will be presented. 

In the case of ferromagnetic coupling ($J < 0$) the 
ground state energy is 
$E_{0, \: J < 0} = - \vert J \vert / 4 - t$ with wave function
$\vert \psi_{0} \rangle_{J < 0} \equiv 
\vert \psi_{\rm DE} \rangle_{J < 0, \: z} = 
\vert \Uparrow_{z} \uparrow_{z}, \:  \Uparrow_{z} 0 \rangle + 
\vert \Uparrow_{z} 0, \: \Uparrow_{z} \uparrow_{z} \rangle$,
where $\Uparrow_{z}$ and $\uparrow_{z}$ refers to the $z$ 
component of the impurity and conduction electron spins, 
respectively. Ferromagnetism arises here via an Ising type 
coupling, which allows for description of the ground state 
within a simple semiclassical approximation.\cite{detheory} 
For $J > 0$ the situation is completely changed due to the 
singlet formation of local and conduction electron spins. 
This causes a mixing of the total spin and an enhancement 
of the Hilbert space, where now 16 elements have to be 
considered. The ground state energy is given by 
$E_{0, \: J > 0} = - J / 4 - 
\sqrt{J^{2} + 2 J t + 4 t^{2}} / 2$ with wave functions
$\vert \psi_{0} \rangle_{J > 0} \propto \vert \psi_{\rm KS} 
\rangle_{z} + [ 1 / (J/4 - E_{0, \: J > 0}) ] \: 
\{ \vert \Uparrow_{z} \downarrow_{z}, \: \Uparrow_{z} 0 \rangle 
+ \vert \Uparrow_{z} 0, \: \Uparrow_{z} \downarrow_{z} \rangle
- \vert \Uparrow_{z} \uparrow_{z}, \: \Downarrow_{z} 0 \rangle 
- \vert \Downarrow_{z} 0, \: \Uparrow_{z} \uparrow_{z} \rangle \}$,
where the Kondo singlet $\vert \psi_{\rm KS} \rangle_{z}$ states are
$ \vert \Uparrow_{z} \downarrow_{z}, \: \Uparrow_{z}, 0 \rangle 
- \vert \Downarrow_{z} \uparrow_{z}, \: \Uparrow_{z} 0 \rangle
+ \vert \Uparrow_{z} 0, \: \Uparrow_{z} \downarrow_{z} \rangle
- \vert \Uparrow_{z} 0, \: \Downarrow_{z} \uparrow_{z} \rangle$. 
$\vert \psi_{0} \rangle_{J > 0}$ involves six basis elements 
(the degeneracy is partially lifted by conduction electron hopping) 
and hence falls outside the four dimensional space needed to 
establish DE for $J < 0$. In order to invoke DE as well, all 
three spin directions, $x$, $y$, and $z$, have to be considered: 
$\vert \psi_{0} \rangle_{J > 0} \propto 
[1 - 1 / (J/4 - E_{0, \: J > 0}) ] \: \vert \psi_{\rm KS} \rangle_{z} 
+ [ 1 / (J/4 - E_{0, \: J > 0}) ] \: 
\{ \vert \psi_{\rm DE} \rangle_{J > 0, \: x} 
+ \vert \psi_{\rm DE} \rangle_{J > 0, \: y}
+ \vert \psi_{\rm DE} \rangle_{J > 0, \: z} \}$, where
$\vert \psi_{\rm DE} \rangle_{J > 0, \: \alpha = x \: {\rm or} \: y} = 
\{ \vert \Uparrow_{\alpha} \downarrow_{\alpha}, \:  \Uparrow_{\alpha} 0 \rangle 
+ \vert \Uparrow_{\alpha} 0, \:  \Uparrow_{\alpha} \downarrow_{\alpha} \rangle
+ \vert \Downarrow_{\alpha} \uparrow_{\alpha}, \: \Downarrow_{\alpha} 0 \rangle
+ \vert \Downarrow_{\alpha} 0, \: \Downarrow_{\alpha} \uparrow_{\alpha} \rangle
\} / {\sqrt{2}}$ and 
$\vert \psi_{\rm DE} \rangle_{J > 0, \: \alpha = z} = 
  \vert \Uparrow_{z} \downarrow_{z}, \: \Uparrow_{z} 0 \rangle
+ \vert \Uparrow_{z} 0, \: \Uparrow_{z} \downarrow_{z} \rangle$. 
In spite of this extra complication, it is apparent from the above 
that in both cases, $J < 0$ and $J > 0$, spin polarons are formed. 

Going beyond the two site approximation, spin polaron formation
can be derived directly from the KLM Hamiltonian, which 
can be written as:
$H \approx -t \sum_{i} ( c^{\dagger}_{i, \sigma}
c^{}_{i+1, \sigma} + {\rm h.c.} ) + J/2 \sum_{i} 
(n_{i, \uparrow} - n_{i, \downarrow}) S^{z}_{{\rm d}, j}$. 
Here spin-flip interactions are neglected, since these require 
a much higher energy and are consequently unlikely to be of 
importance to our results. This simplified Hamiltonian, as 
compared to Hamiltonian Eq.\ (\ref{one}), can be solved when the 
electronic wave functions  are treated within the continuum 
approximation and in the limit $N_{c} / N \ll 1$, a case which is 
relevant to small doping concentrations in CMR materials. 
The electronic wave functions, $\psi_{\sigma}(x)$, satisfy a 
standard nonlinear Schr{\"o}dinger equation:  
$\partial^{2}_{x} \psi_{\sigma}(x) + (J m_{{\rm el.}} / 2) 
\vert \psi_{\sigma}(x) \vert^{2} \psi_{\sigma}(x)
= 2 m_{{\rm el.}} E \psi_{\sigma}(x)$
($m_{{\rm el.}}$ being the bare electron mass) 
with soliton solutions $\psi_{\sigma}(x) \propto  e^{i x} 
{\rm sech} ( x \sqrt{J m_{{\rm el.}}  / 4} )$. 

These soliton solutions correspond to spin domain walls of 
finite size (kink-antikink pairs) and lead to a gain 
in electronic energy of $- \sigma$ for antiferromagnetic coupling, 
and of $+ \sigma$  for the ferromagnetic case. Physically 
the solutions resemble the dressing of the electron by a finite 
range of parallel (antiparallel) local spins and consequently 
represent polaronic type objects. From the previous considerations 
it can also be concluded that, when including the interactions 
with the phonons, the tendency towards charge localization is 
enhanced and increases this polaronic effect. Since the lattice 
also experiences a renormalization due to the coupling to 
the electronic degrees of freedom, substantial ionic 
displacement patterns will develop and the formation 
of magnetoelastic polarons takes place. Similar results 
are obtained by decoupling electronic and phononic degrees 
of freedom through a homogeneous Lang-Firsov transformation, 
where the localization stems from band narrowing. The 
localization width (polaron radius) is characterized by 
a length scale proportional to $1 / {\sqrt{J}}$ . 
This new length scale differs from the free 
conduction electrons mean free path 
and gives rise to competing time scales: slow motion of 
the polaronic carriers and fast motion of the free electrons 
thus inferring dynamics of two types of particles and a 
close analogy to a two fluid scenario. Since the polarons
are in general randomly 
distributed within the local spin array, these states 
can be viewed as intrinsic inhomogeneities involving 
spin fluctuations and short-range spin correlations. 
In addition these new slow dynamics will exhibit a peak 
in the spin structure factor at $2 k_F - \pi$ instead of 
the simple $2 k_F$ RKKY signal. A similar observation 
has also been made \cite{numerics} using numerical approaches. 

In order to investigate rigorously the ordering of the local 
spins due to the formation of polarons, we first apply,   
an infinite (to avoid truncation errors) unitary transformation, 
${\hat{\rm S}}$, to the bosonized Hamiltonian, Eq.\ (\ref{two}). 
The most effective form of ${\hat{\rm S}}$ is given by: 
${{\hat{\rm S}}} = i ( J / 2 \pi ) {\sqrt{v_{F} / v^{3}_{\sigma}}} 
\sum_{j} \theta_{\sigma}(j) \: S^{z}_{{\rm d}, j}$, which couples 
the conduction electron spins directly to the localized spins. 
Secondly, we explicitly take into account the Luttinger 
liquid character of the Bose fields, i.e., use their
non-interacting expectation values such that the 
effective Hamiltonian for the local spins is derived as:
\begin{eqnarray}
H_{{\rm eff}} \: &=& \: - {\frac{J^2 v^{2}_{\sigma}}{4 \pi^2 v_{F}}}
\sum_{j, j^{\prime}} \: \int^{\infty}_{0} dk 
\cos [ k (j - j^{\prime})] \Lambda^{2} (k) \: 
S^{z}_{{\rm d}, j} S^{z}_{{\rm d}, j^{\prime}}
\nonumber \\
&+& \: {\frac{J}{2 \pi \lambda}} \sum_{j}
\{ \cos[K(j)] + \cos[2 k_{F} j] \} S^{x}_{{\rm d}, j}
\nonumber \\
&-& \: {\frac{J}{2 \pi \lambda}} \sum_{j}
\sin[K(j)] \sin[2 k_{F} j] S^{z}_{{\rm d}, j} \; .
\label{heff}
\end{eqnarray}
Here $K(j)$ stems from the unitary transformation and counts all 
the $S^{z}_{{\rm d}, j}$'s to the right of the site $j$ 
and subtracts all those to the left of $j$: $K(j) = 
(J / 2 v_{F}) \sum_{l = 1}^{\infty} ( S^{z}_{{\rm d}, j - l} - 
S^{z}_{{\rm d}, j - l} )$. This term gives the crucial 
difference between the Kondo lattice and dilute Kondo lattice, 
as will be explained in the following. The most important term 
in Eq.\ (\ref{heff}) is the first one, 
which shows that a ferromagnetic coupling emerges 
even in the dilute Kondo lattice model. This coupling is 
non-negligible for $N_{d} > N_{c}$ and $j - j^{\prime} 
\le \lambda$ and its strength will decrease with the 
distance between impurity spins. 
Thus, $\lambda$ represents the effective delocalization 
length related to the spatial extent of the polarons, i.e., the 
effective range of DE. Thus DE will vanish if the distance between 
the impurity spins is larger than $\lambda$. In general $\lambda$ 
will depend on $J$, $N_{c}$ or even $N_{d}$, but we will use its 
low density value: $\lambda \approx {\sqrt{2 / J}}$. 
Consequently, we approximate it by its nearest neighbor form: 
${\cal I} = (J^2 v^2{\sigma} / 2 \pi^2 v_{F} )
\int^{\infty}_{0} dk \cos k \Lambda^{2} (k)$. 

For the Kondo lattice model, $K(j)$ is vanishingly small as 
the number of $d$-spins to the left and the right of a given 
site $j$ is the same. The effective Hamiltonian can thus be 
replaced by a random transverse field Ising model:\cite{fisher}  
$H_{{\rm eff}} = - {\cal I} \sum_{j} S^{z}_{{\rm d}, j} 
S^{z}_{{\rm d}, j + 1} - \sum_{j} h_{j} S^{x}_{{\rm d}, j}$, 
where the ferromagnetic coupling strictly vanishes if 
$\ell > \lambda$. The random fields, $h_{j}$, are generated 
by $(1 + \cos[2 k_{F} j])$ at large distances, where 
$\cos[2 k_{F} j]$ oscillates unsystematically with respect
to the lattice. The large values $\cos[2 k_{F} j] \approx 1$ 
which are responsible for spin flips, are then widely separated
and are driven by a cosine distribution  similarly to  
spin-glasses.\cite{abrikosov} If we have a small 
concentration of holes in the array of localized 
spins, then - opposite to the previous case - $K(j)$  
is non-vanishing since the hole spins are no longer equally 
distributed to the left and the right of a given site. 
This yields $K(j) \approx (-1)^{j} (J / 2 v_{F})$,
which gives rise to a staggered field and antiferromagnetic 
ordering. 

Since our main interest here is to explore the occurrence of 
ferromagnetism in the presence of the Hubbard and phonon terms, 
we focus on the transition between the paramagnetic and the 
ferromagnetic phase. This is controlled by a critical coupling
$J_{{\rm crit.}} = (\pi / 4) \sin ( \pi n / 2 ) \: 
\{ 1 - U / [2 \pi sin (\pi n / 2)] +  
\alpha^2 / [2 \pi K \sin(\pi n / 2)] \}^{1/2}$. 
For values $J < J_{\rm crit.}$ a paramagnetic state exists 
which is dominated by polaronic fluctuations. For $J > J_{\rm crit.}$
ferromagnetism appears. The transition between these phases is 
of order-disorder type with variable critical exponent $\delta =
J_{\rm crit.} / J$. It can be seen that, in accordance with 
our previous observation, the Hubbard term stabilizes the 
ferromagnetic phase, while the phonons counter this and tend to 
increase the polaronic regime. This paramagnetic polaron 
state can be viewed as a Griffiths phase, since the critical 
exponent is variable and the spin-spin correlation function 
is given by: $(\xi/x)^{5/6} e^{-(3/2)(2\pi^{2} x/\xi)^{1/3}} 
e^{-x/ \xi}$, where $\xi \approx 1/ \delta^2$ is the correlation 
length. At finite temperature the susceptibility in this phase 
is proportional to  $T^{2 \delta -1} (\ln T)^2$, while the 
specific heat follows a $T^{2 \vert \delta \vert}$ dependence. 
This regime can be viewed as a paramagnet with locally ordered 
ferromagnetic regions, again manifesting the analogy to a two 
fluid picture. 

In summary, we have derived an effective Hamiltonian from a 
one-dimensional Kondo lattice model extended to include effects 
stemming fom the lattice and in the presence of an onsite 
Hubbard term, which accounts for the conduction electron 
Coulomb repulsion. The results are: i) A ferromagnetic 
phase appears at intermediate $\vert J \vert$ due to 
forward scattering by delocalized conduction electrons. 
ii) Ferromagnetism is favoured by the Hubbard term, 
while it is suppressed by the electron-phonon coupling. 
iii) The paramagnetic phase is characterized by 
the coexistence of polaronic regimes with intrinsic 
ferromagnetic order and ordinary conduction electrons. 
iv) In the paramagnetic phase, two time scales compete 
with each other - reminiscent of a two-fluid model -  
and the variability of the critical exponents suggests 
the existence of a Griffiths phase. The results are related 
to the small-doping regime of CMR materials which are 
ferromagnets at low temperatures, since here the coupling 
to the phonons has been shown to dominate the 
paramagnetic-ferromagnetic phase transition. 

It is interesting to note the discrepancy between 
infinite dimensional calculations and the present one dimensional 
result. Many calculations to model CMR \cite{phonontheory} 
have been made in dynamical mean-field theory, 
which is an infinite dimensional 
approximation and therefore incapable of capturing spatial 
inhomogeneities. In the present work we approach the CMR 
materials via a one dimensional approximation, but with 
techniques able to describe fluctuations of short-range order. 
Our results show that strong intrinsic spatial inhomogeneities of 
Griffiths type dominate the behaviour of the 
Kondo lattice. Consequently the inhomogeneities exhibit clear 
statistical scaling properties as a function of the proximity
to a {\sl quantum} (order-disorder) {\sl critical point}. 
The phonons enhance the inhomogeneities, which in a good 
approximation behave as a supercritical 
(metastable) phase of a two fluid model.

Even though various bosonization schemes have been used for 
the one-dimensional KLM,\cite{boso,graeme} non of the previous 
approaches took into account phonons and the possibility of diluting
the array of impurity spins. The inclusion of phonon degrees of 
freedom has been shown to be relevant in creating
local magnetic inhomogeneities. It is important to
mention that the properties of the system are 
driven by intrinsic inhomogeneities. This means that, in a 
renormalization group approach, the dimensionality should not 
matter.\cite{fisher} Thus, similar behaviour is expected in 
realistic two- and three-dimensions, which clerly merit further 
detailed study.

\end{document}